\documentclass[conference]{template/IEEEtran}
\IEEEoverridecommandlockouts

\usepackage{times}
\usepackage{soul}
\usepackage{url}
\usepackage[hidelinks]{hyperref}
\usepackage[utf8]{inputenc}
\usepackage[small]{caption}
\usepackage{enumitem}
\usepackage{graphicx}
\usepackage{amsmath}
\usepackage{amsthm}
\usepackage{booktabs}
\usepackage{algorithm}
\usepackage{algorithmic}
\usepackage[switch]{lineno}
\usepackage{lipsum}
\usepackage{svg}
\usepackage{subcaption}
\usepackage{booktabs}
\usepackage{array}
\usepackage{makecell}
\usepackage{graphicx} 
\usepackage{mathtools}
\usepackage{enumitem}

\usepackage{xurl}
\usepackage{mdframed}
\usepackage{listings}
\usepackage{xcolor} 

\lstdefinelanguage{json}{
  basicstyle=\ttfamily\small,
  string=[s]{"}{"},
  comment=[l]{//},
  morecomment=[s]{/*}{*/},
  showstringspaces=false,
  breaklines=true,
  literate=
   *{0}{{{\color{blue}0}}}{1}
    {1}{{{\color{blue}1}}}{1}
    {2}{{{\color{blue}2}}}{1}
    {3}{{{\color{blue}3}}}{1}
    {4}{{{\color{blue}4}}}{1}
    {5}{{{\color{blue}5}}}{1}
    {6}{{{\color{blue}6}}}{1}
    {7}{{{\color{blue}7}}}{1}
    {8}{{{\color{blue}8}}}{1}
    {9}{{{\color{blue}9}}}{1}
    {:}{{{\color{red}:}}}{1}
    {,}{{{\color{red},}}}{1}
    {\{}{{{\color{red}\{}}}{1}
    {\}}{{{\color{red}\}}}}{1}
    {[}{{{\color{red}[}}}{1}
    {]}{{{\color{red}]}}}{1},
}

\usepackage{latexsym}
\usepackage[most]{tcolorbox}
\usepackage{listings}

\definecolor{headerblue}{RGB}{0, 20, 140}
\definecolor{taskgold}{RGB}{184, 175, 80}
\definecolor{refgreen}{RGB}{50, 120, 40}
\definecolor{unconblue}{RGB}{30, 40, 150}
\definecolor{contamred}{RGB}{170, 40, 30}

\tcbset{
    codebox/.style={
        colback=white,
        colframe=gray!30,
        fontupper=\ttfamily\small,
        sharp corners=false,
        arc=2pt,
        boxrule=1pt,
        left=5pt,
        right=5pt,
        enhanced,
        breakable,
        verbatim
    }
}
\newtcolorbox{contentbox}[2]{
    colback=#1!5!white,
    colframe=#1,
    title=#2,
    fonttitle=\bfseries\large,
    coltitle=white,
    enhanced,
    attach boxed title to top left={xshift=5mm, yshift=-3mm},
    boxed title style={colback=#1, sharp corners=false, arc=3pt},
    sharp corners=false,
    arc=10pt,
    boxrule=2pt,
    top=5mm
}

\usepackage{cite}
\usepackage{amsmath,amssymb,amsfonts}
\usepackage{algorithmic}
\usepackage{graphicx}
\usepackage{textcomp}
\usepackage{xcolor}
\def\BibTeX{{\rm B\kern-.05em{\sc i\kern-.025em b}\kern-.08em
    T\kern-.1667em\lower.7ex\hbox{E}\kern-.125emX}}
\begin{document}

\title{RAVEN: Retrieval-Augmented Vulnerability Exploration Network for Memory Corruption Analysis in User Code and Binary Programs}

\author{
  \IEEEauthorblockN{
    Parteek Jamwal$^{2,*}$,\;
    Minghao Shao$^{1,2,*}$,\;
    Boyuan Chen$^{1,2,*}$,\;
    Achyuta Muthuvelan$^{2}$\;
    Asini Subanya$^{2}$\\
    Boubacar Ballo$^{2}$\
    Kashish Satija$^{2}$\;
    Mariam Shafey$^{2}$\;
    Mohamed Mahmoud$^{2}$\;
    Moncif Dahaji Bouffi$^{2}$\\
    Pasindu Wickramasinghe$^{2}$\;
    Siyona Goel$^{2}$\
    Yaakulya Sabbani$^{2}$\;
    Hakim Hacid$^{3}$\;
    Mthandazo Ndhlovu$^{3}$\\
    Eleanna Kafeza$^{3}$\;
    Sanjay Rawat$^{3}$\;
    Muhammad Shafique$^{2}$
    \thanks{$^{*}$Authors contributed equally to this work.}
  }
  \IEEEauthorblockA{
    $^{1}${New York University, USA}\;
    $^{2}${New York University Abu Dhabi, UAE}\;
    $^{3}${Technology Innovation Institute, UAE}\;
  }
}

\maketitle

\begin{abstract}
Large Language Models (LLMs) have demonstrated remarkable capabilities across various cybersecurity tasks, including vulnerability classification, detection, and patching. However, their potential in automated vulnerability report documentation and analysis remains underexplored. We present RAVEN (Retrieval Augmented Vulnerability Exploration Network), a framework leveraging LLM agents and Retrieval Augmented Generation (RAG) to synthesize comprehensive vulnerability analysis reports. Given vulnerable source code, RAVEN generates reports following the Google Project Zero Root Cause Analysis template. The framework uses four modules: an Explorer agent for vulnerability identification, a RAG engine retrieving relevant knowledge from curated databases including Google Project Zero reports and CWE entries, an Analyst agent for impact and exploitation assessment, and a Reporter agent for structured report generation. To ensure quality, RAVEN includes a task specific LLM Judge evaluating reports across structural integrity, ground truth alignment, code reasoning quality, and remediation quality. We evaluate RAVEN on 105 vulnerable code samples covering 15 CWE types from the NIST-SARD dataset. Results show an average quality score of 54.21\%, supporting the effectiveness of our approach for automated vulnerability documentation.
\end{abstract}

\section{Introduction}
Large Language Models (LLMs) have demonstrated transformative capabilities across a wide spectrum of domains, reshaping how we approach natural language processing, code generation, and reasoning \cite{shao2024survey}. In cybersecurity, LLMs have been applied to vulnerability detection in source code \cite{xi2025trace}, penetration testing automation \cite{gelei2024pentestgpt}, automated program repair  \cite{kulsum2024vrpilot} and agent-based attack/defense for simulating real-world cybersecurity scenarios \cite{shao2024empirical}. These efforts highlight the growing integration of LLMs into security workflows.

Despite these advances, the application of LLMs in generating comprehensive vulnerability analysis reports remains underexplored. Professional security organizations such as Google produce detailed reports documenting root cause analyses, exploitation techniques, and remediation strategies \cite{zeroRootCauseAnalyses}. These reports follow structured templates capturing key information such as vulnerability summaries, attack surfaces, exploit primitives, and patch guidance. However, automating the synthesis of such reports using LLMs introduces challenges that have not been adequately addressed.

Automated vulnerability report generation presents three core challenges. First, it requires integrating multiple subtasks, including vulnerability classification, root cause analysis, exploitation assessment, and patch synthesis, into a single coherent report, which is more complex than solving each subtask in isolation. Second, vulnerability analysis often depends on extended code contexts, call chains, and cross-file dependencies, demanding robust long-context reasoning; yet LLMs can suffer from the ``lost in the middle'' phenomenon when critical evidence appears in the middle of long inputs. Third, evaluating generated reports is inherently multi-dimensional, for example factual accuracy, technical depth, and remediation validity, and building automated evaluation frameworks for such domain-specific outputs remains an open problem.

\begin{figure}
  \centering
  \includegraphics[width=\columnwidth]{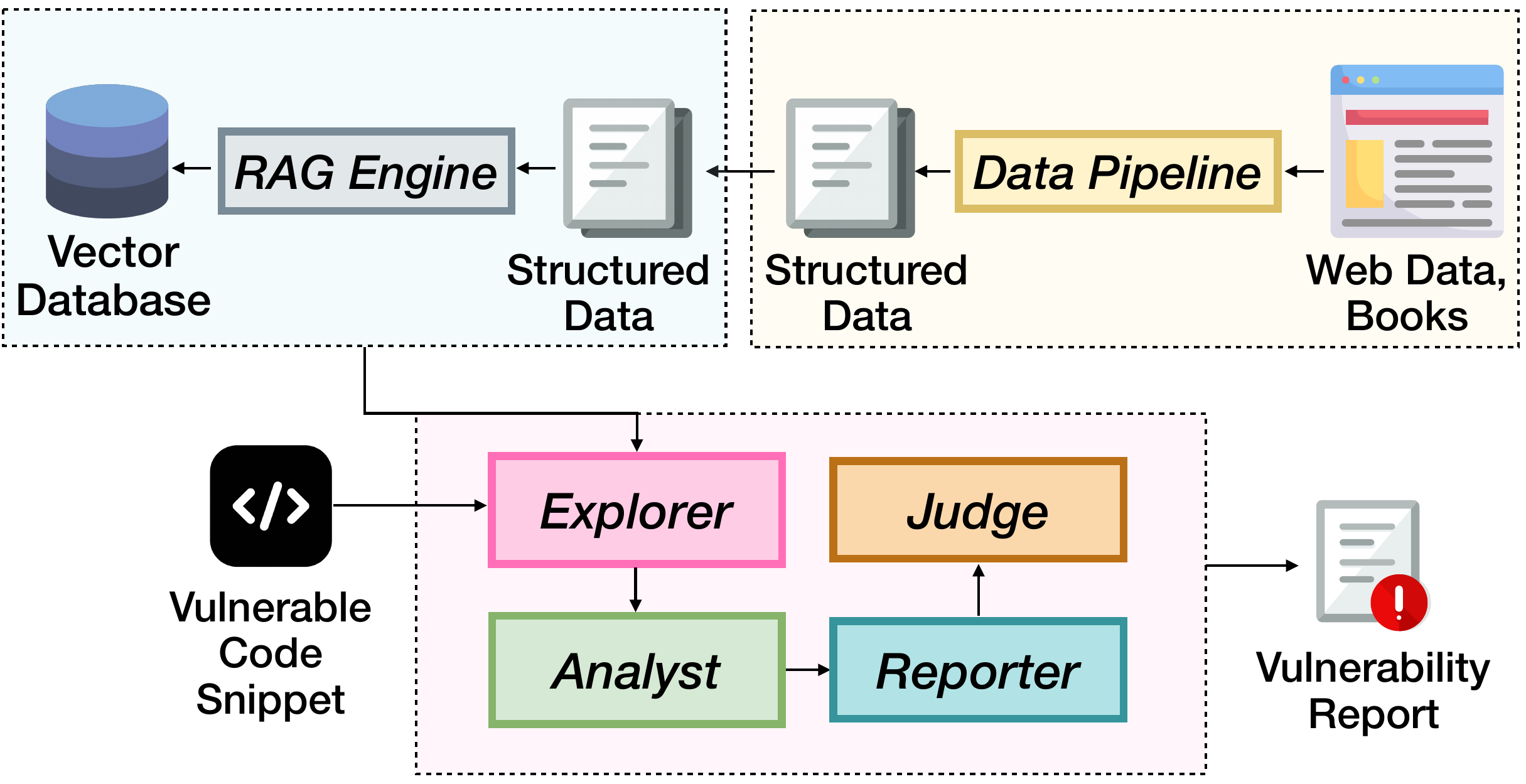}
  \caption{Architectural Overview of RAVEN. (1) a \textbf{Data Collection Pipeline} that transforms raw web pages and PDFs into a structured format (2) a \textbf{RAG Engine} that indexes the structured data into a vector database for retrieval (3) an \textbf{Agentic System} that takes in the vulnerable code snippet and generates a comprehensive, Google Project Zero Style Vulnerability Report.}
\end{figure}

To address these challenges, we propose RAVEN (Retrieval Augmented Vulnerability Exploration Network), a multi agent framework that leverages LLMs and Retrieval Augmented Generation (RAG) to automatically synthesize comprehensive vulnerability analysis reports. RAVEN implements a four phase agentic workflow comprising an Explorer agent for initial vulnerability identification and CWE classification, a RAG engine that retrieves relevant vulnerability knowledge from curated databases including Google Project Zero reports and MITRE CWE entries, an Analyst agent for in depth impact and exploitation assessment, and a Reporter agent that generates structured reports following the Google Project Zero Root Cause Analysis template. To ensure report quality, RAVEN incorporates a task specific LLM as a Judge module that evaluates generated reports across four dimensions: structural integrity, ground truth alignment, code reasoning quality, and remediation quality. 

Overall, our contributions can be summarized as follows:
\begin{enumerate}[leftmargin=*]
\item We propose RAVEN, a novel multi-agent framework that integrates RAG with LLM agents to automate the synthesis of professional-grade vulnerability analysis reports.
\item We design a comprehensive evaluation framework utilizing LLM-as-a-Judge methodology to assess vulnerability reports across multiple quality dimensions.
\item We conduct extensive experiments on the NIST-SARD dataset, demonstrating RAVEN's effectiveness in generating accurate and comprehensive vulnerability reports across diverse CWE categories.
\end{enumerate}

\section{Related Works}

\textbf{LLM-based Vulnerability Detection for Code.}
LLMs have shown promising capabilities in identifying source code security vulnerabilities. Early approaches leveraged encoder-only models such as CodeBERT \cite{feng-etal-2020-codebert} and GraphCodeBERT \cite{guo2021graphcodebert} for vulnerability classification tasks, while recent work has shifted toward decoder-only architectures including GPT-4 and CodeLlama~\cite{rozière2024codellamaopenfoundation}. Studies demonstrate LLMs can perform vulnerability detection through prompt engineering and chain-of-thought reasoning~\cite{answar2025gosonar}, though comprehensive evaluations reveal that current models still struggle with reliable identification and reasoning about security flaws, particularly in complex, multi-file contexts~\cite{li2025svtrustevalc, yildiz-etal-2025-benchmarking}. Multi-model collaboration approaches such as M2CVD~\cite{wang2025m2cvd} and mixture-of-experts frameworks~\cite{lekssays2025llmcpg} have been proposed to enhance detection accuracy across diverse CWE categories.

\textbf{LLM-based Program Repair/Fix.}
Automated program repair using LLMs has emerged as an active research area. VRpilot~\cite{kulsum2024vrpilot} introduced chain-of-thought prompting combined with iterative patch validation feedback, demonstrating improvements over baseline techniques for vulnerability repair in C and Java. VulRepair~\cite{fu2022vulrepair} applied T5-based models for automated software vulnerability repair, while subsequent work explored few-shot learning  ~\cite{feng2024prompting} and fine-tuning strategies~\cite{zhou2024vulmaster}. Recent studies have also investigated the integration of static analyzers with LLMs to improve patch quality and reduce false positives~\cite{li2025iris}. However, challenges remain in generating patches that correctly address root causes without breaking functionality, particularly for complex vulnerabilities tied to project-specific design patterns~\cite{pearce2023iris}.

\textbf{LLM Agents for Security Tasks.}
Multi-agent LLM systems have been developed to automate complex security workflows. PentestGPT~\cite{gelei2024pentestgpt} pioneered modular task decomposition for penetration testing, utilizing a pentesting task tree for structured reasoning. PentestAgent~\cite{shen2025pentestagent} extended this paradigm by incorporating Retrieval-Augmented Generation (RAG) to enhance domain knowledge \cite{shao2025craken} and automate intelligence gathering, vulnerability analysis, and exploitation stages. Related multi-agent approaches have also been explored in adjacent domains, including hardware design~\cite{xiao2025trojanloc}, offensive security~\cite{udeshi2025d}, and vulnerability detection~\cite{xi2025trace}. These frameworks leverage RAG to retrieve relevant vulnerability information from curated databases, mitigating the limitations of LLM context windows and outdated training knowledge. Similar multi-agent architectures have been applied to red teaming and defensive security applications~\cite{he2025readteaming}.

\textbf{LLM-as-a-Judge for Evaluation.}
Evaluating the quality of LLM-generated security analyses presents unique challenges due to the domain-specific nature of vulnerability reports. The LLM-as-a-Judge paradigm~\cite{li2025llmasajudge} offers a scalable alternative to manual annotation by employing LLMs to assess outputs across multiple dimensions. This approach has been adopted for evaluating code generation quality, vulnerability detection rationales, patch correctness~\cite{mou2025canyoutrust} and agentic tasks ~\cite{shao2025towards, chen2025metacipher}. While LLM judges demonstrate strong agreement with human evaluators on well-defined criteria, concerns remain regarding their robustness to adversarial manipulations and their reliability in out-of-domain scenarios~\cite{raina2024isllmasajudgerobust}.

Our work differs from prior approaches by integrating these components into a unified framework for comprehensive vulnerability report generation, combining multi-agent collaboration with RAG-enhanced retrieval and systematic LLM-based evaluation across multiple quality dimensions.

\section{Methodology}

RAVEN’s architecture consists of three major technical components, each addressing a distinct challenge in automated vulnerability analysis. These are (i). Data Collection Pipeline (ii). RAG Engine (iii). Agentic Framework. All these modules are discussed in detail in the subsequent sections.

\subsection{Data Collection Pipeline}

In this section, we'll describe our approach for automated data collection techniques from webpages and books. 

\begin{figure}
  \centering
   \includegraphics[width=\columnwidth]{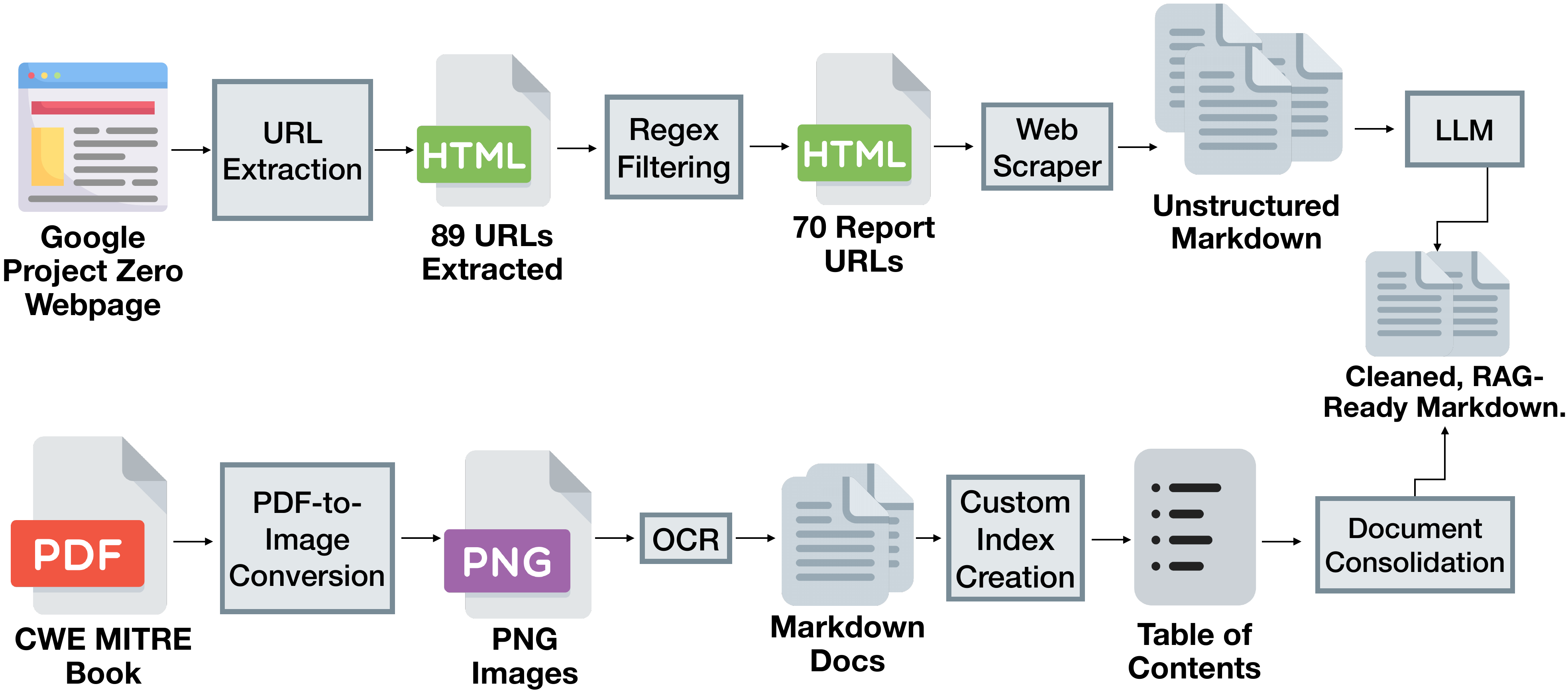}
  \caption{Data Collection Pipeline. (\textbf{Top}) URL extraction and web scraping of security reports; (\textbf{Bottom}) PDF-to-markdown conversion using OCR and Table of Contents (TOC) indexing. The final output is a cleaned, structured markdown for RAG.}
\end{figure}

\subsubsection{Google Project Zero}
Through an extensive analysis of publicly available detailed vulnerability reports, we found Google Project Zero’s 0-days-in-the-wild repository \cite{zeroRootCauseAnalyses} consisting of 70 reports detailing the essential reasoning behind various 0-day exploits. 

To extract the content from the reports, we first retrieve all URLs from the Google Project Zero homepage using the Crawl4AI library \cite{crawl4ai2024}. A BFS (Breadth-First Search) deep crawling strategy with a depth level of 2 is used. This crawling process yields \textbf{89} unique same-domain URLs. On manual inspection of these URLs, we noticed that some of these links point to non-report sources. Therefore, to retain links that only contain the reports, we define a custom regex filter. Using this filter, we go from \textbf{89} to \textbf{70} valid CVE report links which are saved for report content extraction.

\begin{mdframed}
\url{https://googleprojectzero.github.io/0days-in-the-wild/0day-RCAs/YYYY/CVE-YYYY-########.html}
\end{mdframed}

Similarly to the URL extraction phase, a BFS deep-crawling strategy with a depth of 1 is used to ensure that all content from the target  URL and the embedded links within the original report, such as cited advisories, GitHub repositories, and other relevant links, are captured. A list of HTML tags and CSS selectors (e.g., navigation bar, footer, etc.) to avoid is also specified. All embedded URL content is encapsulated within structural tags: \texttt{<START OF CONTENT FOR [URL]>}  and \texttt{<END OF CONTENT FOR [URL]>}. To manage redundancy, the embedded URL content is limited to three appearances per report; subsequent instances of the same URL are skipped to prevent unnecessary context window inflation. The output of this process is \textbf{70} markdown files, each containing the raw text of the vulnerability reports.


Lastly, we perform an LLM-based cleaning of the raw reports to transform them into structured, RAG-ready vulnerability documents. All these files are passed to an LLM along with a specialized $\approx$ 6200 token prompt that guides it on how to parse, clean, and format the raw scraped content into a consistent schema optimized for Retrieval-Augmented Generation (RAG).

\subsubsection{CWE MITRE Book}
In addition to Google Project Zero vulnerability reports, we also included a 2,735-page PDF released by MITRE \cite{cwe417}, containing 1,321 CWE entries, in our list of documents to ingest into the RAG engine. To transform this PDF into a RAG-ready knowledge base, we implemented the following three-stage pipeline
\begin{enumerate}[leftmargin=*]
    \item \textbf{PDF to Image Conversion}: Every page of the PDF was exported as an individual PNG file using the \texttt{pdf2image} library, resulting in 2,735 PNG images.
    \item \textbf{OCR Processing}: Each saved image is then processed by Chandra \cite{chandra2025}, a highly accurate OCR model that converts images and PDFs into structured markdown while preserving layout information. Since the PDF contains tabular data, code blocks, and complex data layouts, this model was perfectly suited for high-fidelity text extraction.

     \item \textbf{Knowledge Base Consolidation}: At this stage, we have 2,735 markdown documents, one for each page of the PDF. Using the markdown documents for the Table of Contents from the book, we task the LLM to generate an index containing the \textbf{Chapter Name}, \textbf{Start page}, and \textbf{End page}. The index is then used to create chapter specific documents.
\end{enumerate}

\subsection{RAG Engine}

\begin{figure}
  \centering
 \includegraphics[width=\columnwidth]{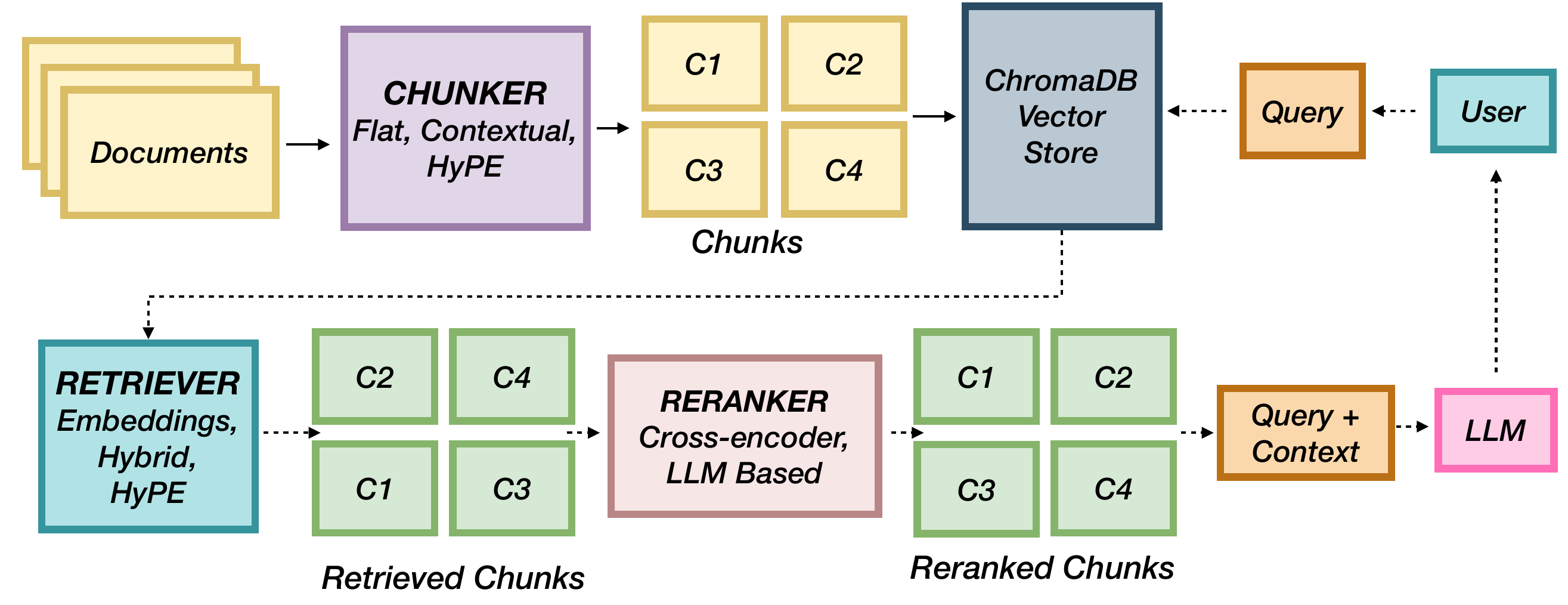}
  \caption{Overview of RAVEN's RAG Engine. It comprises \textbf{3} chunking strategies (Flat, Contextual, HyPE), \textbf{3} retrieval methods (Embeddings, Hybrid, HyPE) and \textbf{2} rerankers (Cross-encoder, LLM-based).}
\end{figure}

RAVEN's RAG engine implements a modular architecture comprising three chunking, three retrieval, and two ranking components, all built upon ChromaDB as the persistent vector database with HNSW (Hierarchical Navigable Small World) indexing for efficient approximate nearest neighbor search. For document ingestion, we first have a flat chunking approach, which uses a fixed-size sliding window on a given document, thereby splitting it into arbitrary fixed-size windows of text. Next, we have the Contextual Chunking strategy based on Anthropic's Contextual Retrieval \cite{ContextualRetrievalAI} approach. The idea is to prepend an LLM-generated chunk-specific explanatory context in relation to the entire document, usually 50-100 tokens, to each chunk before embedding and saving it to the vector database. Lastly, we have the HyPE (Hypothetical Prompt Embeddings) approach, where 3-5 hypothetical, chunk-specific questions are generated for each document chunk, produced by the flat chunking strategy. These generated hypothetical questions become the chunk's representation. Consequently, the retrieval process for the HyPE strategy shifts from query-to-document to query-to-hypothetical questions. Once the chunks are created, they are converted into embeddings, except for the HyPE approach, where the hypothetical questions are embedded instead of document chunks.

Moving on to the retrieval process, we first implemented a standard embedding-based approach, where we perform a pure vector-based similarity search between the embedded query and the ingested chunks. Building on this, we have the hybrid approach, where we combine semantic (vector) search and keyword (BM25) search scores for enhanced retrieval. We use the following scoring mechanism \begin{equation}
        \text{score}_{\text{final}} = w_{\text{semantic}} \cdot \text{score}_{\text{semantic}} + w_{\text{keyword}} \cdot \text{score}_{\text{keyword}}
    \end{equation}
where the default weights are 0.6 for semantic and 0.4 for keyword components. Lastly, we have the HyPE retrieval strategy that performs query-to-generated-hypothetical question matching.

Once the initial set of chunks is retrieved, it is passed into the Reranker module, which plays a crucial role in refining the results obtained from the initial retrieval process. RAVEN's RAG Engine comprises two reranking modules, namely a Cross-Encoder based reranker where the query and the document are jointly processed to compute the relevance score. Cisco's SecureBERT 2.0 \cite{aghaeiSecureBERT20Advanced2025} which is a domain-specific transformer model optimized for cybersecurity tasks is used as the Cross-Encoder model. In addition to the Cross Encoder, we also have a pointwise LLM-based reranker that scores document relevance on a 0–10 scale.

\begin{figure*}
  \centering
     \includegraphics[width=1.05\textwidth]{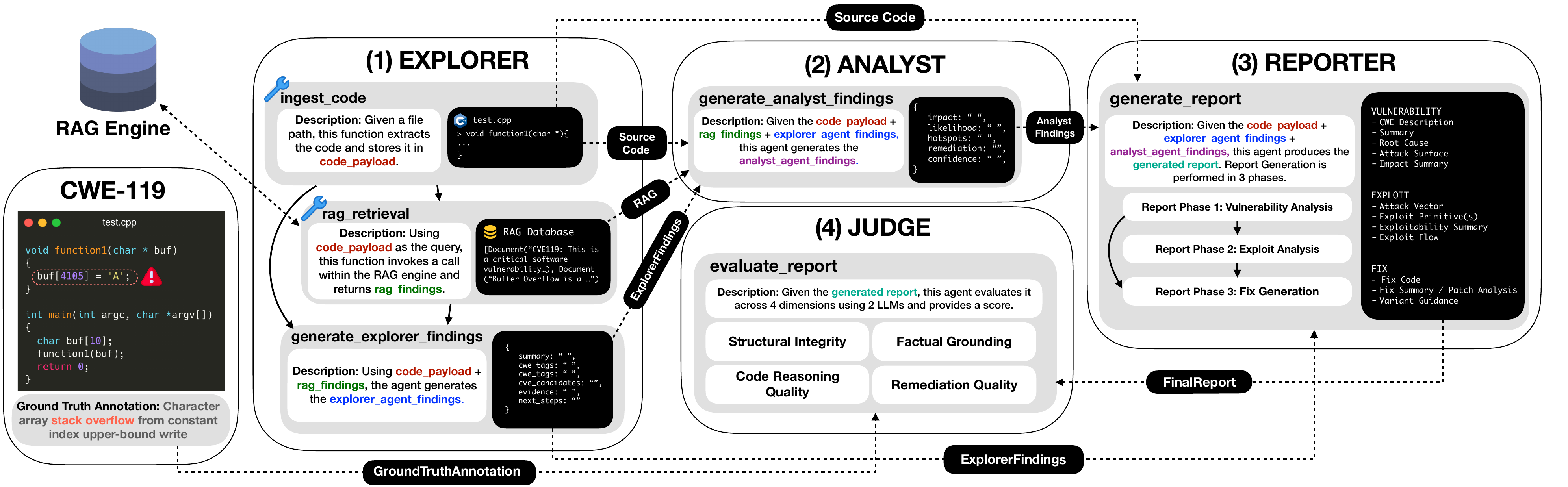}
  \caption{The RAVEN Agentic Pipeline. It orchestrates four specialized agents for end-to-end vulnerability analysis: (1) the \textbf{Explorer} reads the vulnerable code \& queries the RAG engine to generate initial findings; (2) the \textbf{Analyst} refines these findings by assessing impact and likelihood; (3) the \textbf{Reporter} synthesizes a three-phase report (vulnerability, exploit, and fix); and (4) the \textbf{Judge} scores the final output against ground truth across four qualitative dimensions}
\end{figure*}

\subsection{Agentic Workflow}\label{sec:agentic_workflow}

RAVEN implements a four-phase multi-agent system for automated vulnerability analysis and report generation. The workflow comprises four phases: Exploration, Analysis, Report Generation, and Report Evaluation via LLM as a Judge.

The workflow begins by providing the Explorer Agent with a filepath to the vulnerable code snippet. The agent invokes \texttt{\textbf{ingest\_code}} function and extracts all source code, storing it in the \texttt{\textbf{code\_payload}} variable. Once extracted, the explorer agent uses the code snippet as the query and calls \texttt{\textbf{rag\_retrieval}} function, thereby retrieving the most relevant documents from the RAG Engine. The retrieved context and extracted code are fed into \texttt{\textbf{generate\_explorer\_findings}} resulting in the LLM generating an initial set of findings comprising (1) a concise vulnerability class summary, (2) matching Common Weakness Enumeration (CWE) identifiers, (3) Common Vulnerability Enumeration (CVE) identifiers, (4) evidence (line number, snippet, and reason) of where the issue resides, and (5) recommended next steps for immediate mitigation.

Once we have the initial set of explorer agent findings, \texttt{\textbf{ExplorerFindings}}, we use it along with the extracted \texttt{\textbf{code\_payload}} data and retrieved RAG context, and feed them to the analyst agent, which invokes the \texttt{\textbf{analyze}} function. The LLM yields an enhanced set of findings built upon the Explorer Agent's output. These include (1) an impact assessment of the vulnerability, (2) likelihood of exploitation, (3) critical hotspots, (4) confidence level in the analysis, and most importantly, (5) remediation strategies (fixes) for the currently detected vulnerability and  similar variants.

Upon successful generation of the analyst agent’s findings, \texttt{\textbf{AnalystFindings}}, these findings, in conjunction with the code payload, \texttt{\textbf{code\_payload}}, and the explorer agent’s findings, \texttt{\textbf{ExplorerFindings}}, are provided as context to the reporter agent for detailed vulnerability report generation. The report generation happens in three phases:
\begin{enumerate}[leftmargin=*]
    \item \textbf{Vulnerability Analysis}:
    Based on the provided consolidated context, the agent determines whether a vulnerability exists. If it exists, the agent generates a short descriptive title of the detected vulnerability, a 2-4 sentence summary of where the issue appeared, how it was triggered, etc., CWE-based description of the vulnerability, root cause, impact summary, attack surface (potential entry points an attacker could exploit to trigger a vulnerability), and lastly the vulnerable snippet from the original code where the bug resides.
    \item \textbf{Exploit Analysis}: The agent uses Phase 1's findings to produce the attack vector (specific path an attacker takes to execute the exploit), exploit primitives (the low-level capabilities gained upon successful triggering of the vulnerability), exploit steps (a list of 3-5 exploit steps describing how the vulnerability could be triggered), and an exploitability summary. 
    \item \textbf{Fix Generation}: The agent takes the output produced in the earlier two phases and uses it to produce the remediation artifacts. More specifically, the agent produces a code snippet ($<$20 lines) that directly addresses the root cause, an explanation of how/why the fix eliminates the vulnerability, and lastly, the variant guidance, i.e., steps on dealing with future variants of the same exploit.

\end{enumerate}

The output from all three phases is consolidated, fed to a custom markdown renderer to create the final report, and then is evaluated by the Judge agent. The report format is inspired by Google Project Zero's RCA template.

The judge agent uses the original extracted code, ground truth annotation (comes with the original dataset), and the generated report to give its evaluation. This agent utilizes two LLMs, Claude 4.5 Sonnet and Gemini 3.1 Pro each possessing specialized evaluation capabilities to evaluate the generated report across 4 dimensions.

\begin{itemize}[leftmargin=*]
    \item \textbf{Structural Integrity} (0--10): This metric measures how well the generated report adheres to the provided guidelines and the completeness of mandatory fields.
    \item \textbf{Ground Truth Alignment$/$Factual Grounding} (0--10): This metric evaluates factual correctness strictly against provided annotation and source code. A score of 0 is assigned for hallucinated CWEs or CVEs.
    \item \textbf{Code Reasoning Quality} (0--10): This metric assesses the technical depth, specificity, and logical consistency of the report's reasoning.
    \item \textbf{Remediation Quality} (0--10): This metric measures how valid the proposed fix code is relative to the identified root cause in the ground truth annotation.
\end{itemize}

\section{Experiments}

\subsection{LLM Selection}
As outlined in Section \ref{sec:agentic_workflow}, the agentic system operates in four phases: Exploration, Analysis, Report Generation, and Report Evaluation. For the first three phases, we use the same LLM. All experiments are conducted exclusively on Falcon models, Falcon-H1R-7B \cite{teamFalconH1RPushingReasoning2026a} (\texttt{F16}), Falcon-H1-7B-Instruct (\texttt{F16}), Falcon-H1-34B-Instruct (\texttt{Q8}), and 
Falcon3-10B-Instruct (\texttt{F16}), released by the Technology Innovation Institute (TII).

\subsection{Parameters and Features}
For all Falcon models, a temperature of 0.6 and top-p of 0.95 is used. The RAG engine ingests 70 Google Project Zero reports and 1,321 CWE entries extracted from the 2,735-page CWE MITRE PDF into the vector database. For flat chunking, a chunk size of 2000 characters with an overlap of 200 characters is used. For the contextual chunking approach, the LLM-generated contextual summaries prepended to each chunk are limited to 200 characters. The HyPE strategy generates \textbf{3} hypothetical queries per chunk, each capped at 200 characters. The Grok 4.1-Fast model is used as the LLM-reranker and chunk context generator. All retrieval-based methods are designed to retrieve $\texttt{\textbf{top\_k}}$ candidates. With the reranker enabled, the retriever retrieves $\texttt{\textbf{top\_k}} \times \texttt{\textbf{num\_candidates}}$ documents, reranks them, and augments the query with $\texttt{\textbf{top\_k}}$ ranked documents. The values of $\texttt{\textbf{top\_k}}$ and $\texttt{\textbf{num\_candidates}}$ are set to 10 and 2, respectively. For hybrid retrieval, the semantic and keyword weights are set to 0.6 and 0.4, respectively.
\subsection{Benchmark}

To evaluate how well RAVEN's agentic system is able to identify the vulnerability within a given code snippet and generate a detailed report with the fix, we use the NIST-SARD dataset \cite{SoftwareAssuranceReference2021}. It is a large scale dataset curated by NIST as part of the SAMATE project. It contains approximately 450,000 test cases covering more than 150 classes of weaknesses. Each test case is packaged with source code and a manifest file containing the case metadata used as ground truth to assess the system's factual correctness. The evaluation is performed on \textbf{105} memory focused vulnerable samples from the dataset. The composition of our dataset mixture is shown in Table \ref{tab:cwe-distribution}. 

\subsection{Metrics}
The agentic system is evaluated only using LLM-as-a-Judge scores. The judging metrics are Structural Integrity (0-10), Ground-Truth Alignment (0-10), Code Reasoning Quality (0-10) and Remediation Quality (0-10)

\begin{table}[htbp]
\centering
\caption{Test Set Vulnerability Distribution by CWE Type}
\label{tab:cwe-distribution}
\footnotesize
\begin{tabular}{lp{5.2cm}r}
\toprule
\textbf{CWE} & \textbf{Description} & \textbf{\%} \\
\midrule
CWE-119 & Improper Restriction of Operations & 31.43 \\
CWE-787 & Out-of-bounds Write & 6.67 \\
CWE-476 & NULL Pointer Dereference & 6.67 \\
CWE-457 & Use of Uninitialized Variable & 6.67 \\
CWE-416 & Use After Free & 6.67 \\
CWE-170 & Improper Null Termination & 6.67 \\
CWE-122 & Heap-based Buffer Overflow & 6.67 \\
CWE-121 & Stack-based Buffer Overflow & 6.67 \\
CWE-120 & Buffer Copy without Checking Size of Input & 6.67 \\
CWE-415 & Double Free & 5.71 \\
CWE-401 & Missing Release of Memory & 5.71 \\
CWE-193 & Off-by-one Error & 0.95 \\
CWE-126 & Buffer Over-read & 0.95 \\
CWE-124 & Buffer Underwrite & 0.95 \\
CWE-123 & Write-what-where Condition & 0.95 \\
\bottomrule
\end{tabular}
\end{table}

\section{Results}

\begin{table*}[t!]
  \centering
  \caption{Comparison of Falcon Models Across RAG Architectures (Overall Score) $S_{\text{overall}}$}
  \label{tab:rag_comparison_final}

  \footnotesize 
  \setlength{\tabcolsep}{3pt}         
  \renewcommand{\arraystretch}{1.20} 

  \begin{tabular}{l cccc cccc cc}
    \toprule
    \textbf{Model} & 
    \multicolumn{4}{c}{\textbf{Flat Chunking}} & 
    \multicolumn{4}{c}{\textbf{Contextual Chunking}} & 
    \multicolumn{2}{c}{\textbf{HyPE}} \\
    \cmidrule(lr){2-5}\cmidrule(lr){6-9}\cmidrule(lr){10-11}

    & \makecell[c]{EO\\+ CE} & 
      \makecell[c]{HYB\\+ CE} & 
      \makecell[c]{EO\\+ LLM} & 
      \makecell[c]{HYB\\+ LLM} & 
      \makecell[c]{EO\\+ CE} & 
      \makecell[c]{HYB\\+ CE} & 
      \makecell[c]{EO\\+ LLM} & 
      \makecell[c]{HYB\\+ LLM} & 
      \makecell[c]{HyPE\\+ CE} & 
      \makecell[c]{HyPE\\+ LLM} \\
    \midrule

    Falcon H1 34B Instruct         & 7.45 & 7.59 & 7.43 & 7.54 & \textbf{7.68} & 7.62 & 7.41 & 7.65 & 7.44 & 7.34 \\
    Falcon H1R 7B          & 7.13 & \textbf{7.32} & 7.03 & 7.09 & 6.73 & 7.07 & 7.16 & 7.29 & 6.95 & 6.96 \\
    Falcon H1 7B Instruct & 6.80 & 6.66 & 6.77 & 6.82 & 6.50 & 6.66 & 6.80 & 6.76 & \textbf{6.90} & 6.77 \\
    Falcon 3 10B Instruct    & 5.83 & 5.81 & 5.91 & 5.85 & 5.93 & 5.72 & 6.10 & \textbf{6.11} & 5.78 & 5.90 \\

    \bottomrule
  \end{tabular}

  \vspace{2pt}
  \normalsize\textit{Abbreviations: EO = embeddings-only, HYB = hybrid retrieval, CE = cross-encoder reranker, LLM = LLM reranker.}
\end{table*}

We evaluated 4 Falcon models across 10 RAG configurations, resulting in 40 experiments. Each experiment was performed on 105 samples. For every sample, two independent judges generate scores for four dimensions: Structural Integrity (SI), Factual Grounding (FG), Code Reasoning Quality (CR), and Remediation Quality (RQ).

Let $J_1^{SI}, J_1^{GT}, J_1^{CR}, J_1^{RQ}$ denote the scores assigned by Judge~1, and $J_2^{SI}, J_2^{GT}, J_2^{CR}, J_2^{RQ}$ denote the corresponding scores assigned by Judge~2. The overall model score $S_{\text{overall}}$ is computed as the arithmetic mean of all eight criterion scores and is given by:\[
S_{\text{overall}} = \frac{1}{8} \sum_{i=1}^{2} \left( J_i^{SI} + J_i^{GT} + J_i^{CR} + J_i^{RQ} \right).
\]

\begin{figure}[h!]  
    \centering
    \includegraphics[width=\linewidth]{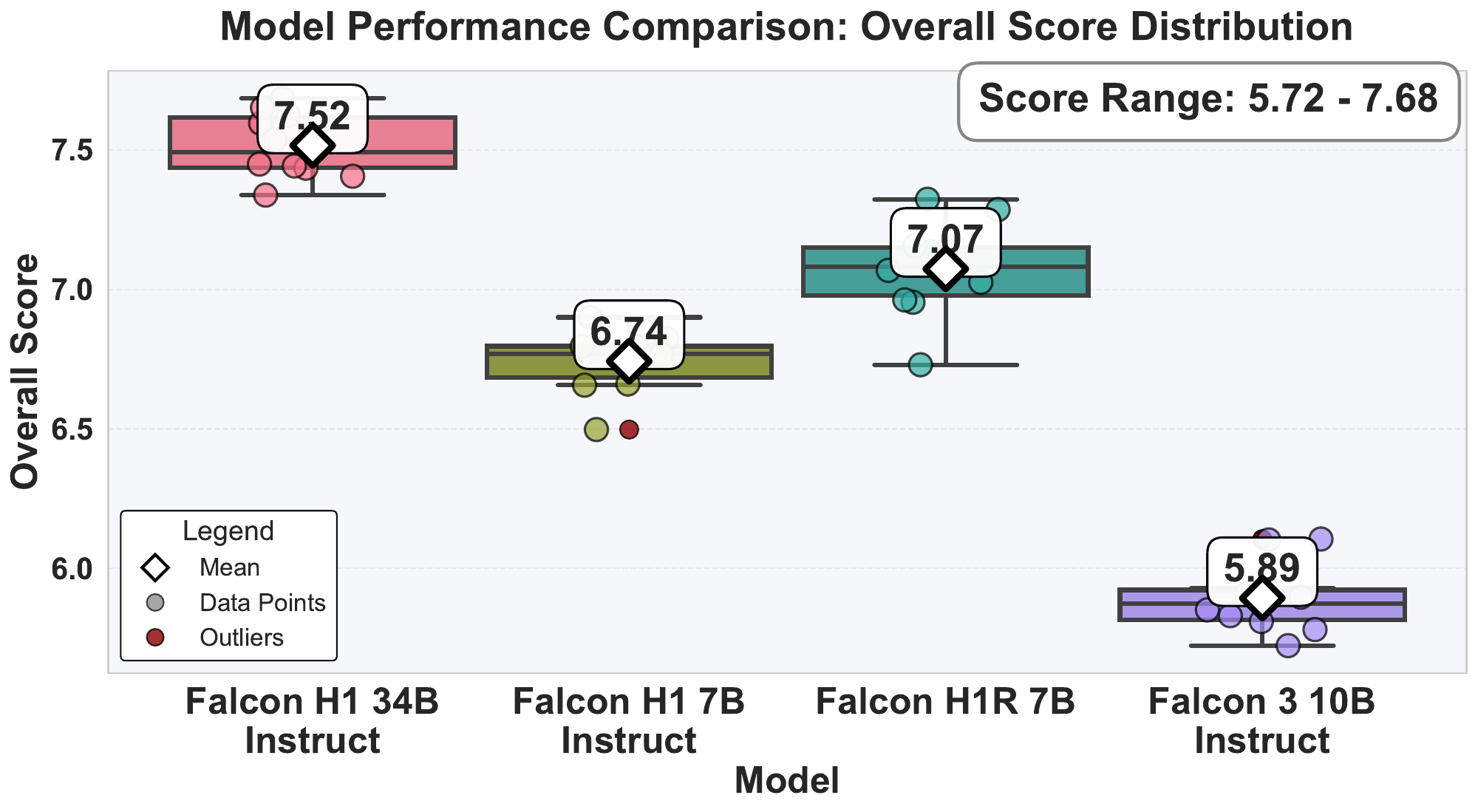} 
    \caption{Box Plot Analysis of Overall Scores for all Falcon Models}
    \label{fig:boxplot}  
    
\end{figure}

Table \ref{tab:rag_comparison_final} shows the overall scores, $S_{\text{overall}}$, of the models. It is clear from the table that there is no specific RAG configuration that outperforms the others.  To gain a deeper insight into the overall scores, we pick a model and analyze how each configuration affects its performance across specific dimensions. To choose the model to inspect, we construct a box plot as depicted in Figure \ref{fig:boxplot} that shows the variance in the model's overall scores. The model with the highest score variance across configurations is picked, thereby allowing us to understand which RAG components influence model performance.

From the box plot in Figure \ref{fig:boxplot}, we can see that the Falcon H1R-7B model exhibits the highest variance compared to the other models. Therefore, we select this model for further analysis of its individual dimensions.

\subsection{Structural Integrity}

\begin{figure*}[h!]  
    \centering
    \includegraphics[width=\linewidth]{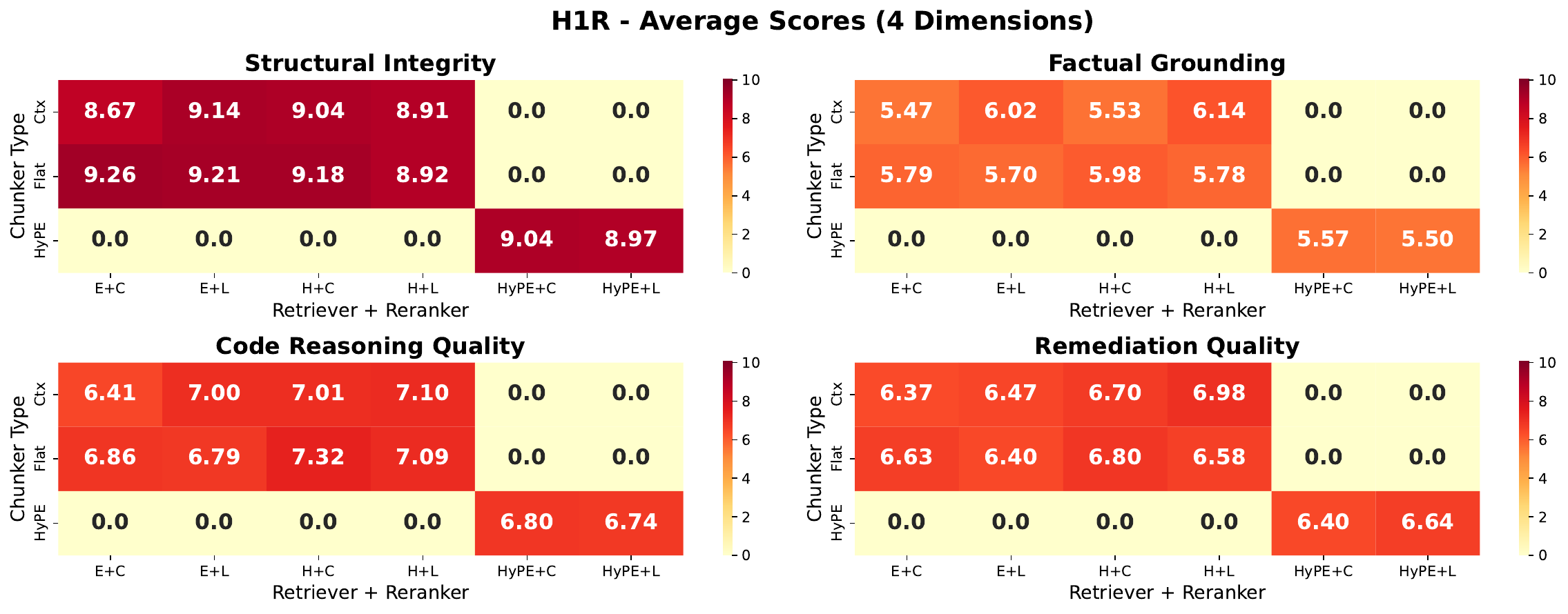} 
    \caption{Falcon H1R-7B Individual Dimension Scores across all RAG Configurations (\normalsize\textit{Abbreviations: E = embeddings-only, H = hybrid retrieval, C = cross-encoder reranker, L = LLM reranker, Ctx=Contextual Chunker})} 
    \label{fig:h1r_dimension_scores}  
\end{figure*}

It is evident from the heatmap shown in Figure~\ref{fig:h1r_dimension_scores} that the Falcon H1R model consistently adhered to the required formatting guidelines across different RAG configurations. Upon examining the judge agent's evaluation logs to identify the causes of point deductions, we found that these were primarily due to two factors:
\begin{itemize}[leftmargin=*]
    \item \textbf{Formatting issues}: Corrupted code rendering, truncated fixes (e.g., missing braces or incomplete function definitions), and the use of literal escape sequences (e.g., \texttt{\textbackslash n}) instead of actual line breaks.
    \item \textbf{False negatives}: Incorrectly classifying a vulnerable code snippet as non-vulnerable, which results in an automatic score of zero for the corresponding evaluation.
\end{itemize}

\subsection{Ground Truth Alignment / Factual Grounding}\label{h1r:fa}

Figure \ref{fig:h1r_dimension_scores} illustrates that the Contextual Chunking + Hybrid Retrieval + LLM Reranker configuration (score=\textbf{6.14}) achieves the highest overall score, closely followed by Contextual Chunking + Embeddings-Based Retrieval + LLM Reranker (score=\textbf{6.02}) and Flat Chunking + Hybrid Retrieval + Cross Encoder Reranker (score=\textbf{5.98}).

The strong performance of Contextual Chunking + Hybrid Retrieval + LLM Reranker configuration can be attributed to the fact that each chunk is accompanied by a semantically rich LLM-generated context, thereby enhancing the chunk's representation. During retrieval, the hybrid strategy combines semantic similarity with keyword-based matching (BM25), enabling the retrieval of a wide range of documents. Given that the input to the RAG engine is a NIST-SARD code snippet, the retrieved set of documents may contain some unrelated/noisy matches due to superficial similarities (shared variable names, function calls, etc.). The initial set of retrievals is then fed to the LLM-based reranker. Given the inherent reasoning capabilities of the LLMs along with the context-enhanced chunk, the reranker is able to generate optimal rankings for the provided query, leading to factually grounded answers (score=\textbf{6.14}). Cross encoders, a class of deterministic rerankers, compute pairwise relevance by calculating the full token-level interactions between query and document. Due to the absence of explicit logical reasoning, the final ranked set of retrievals does not contribute to a high Factual Grounding score (score=\textbf{5.53}). Our empirical findings suggest that additional context confounds the cross encoder's relevance scoring for the H1R model. This performance disparity between the rerankers for contextual chunks can also be viewed by switching the retrieval strategy from hybrid to embeddings-only where an LLM Reranker (score=\textbf{6.02}) outperforms the cross encoder (score=\textbf{5.47}).

Upon analyzing the Flat Chunking + Hybrid Retrieval + Cross Encoder Reranker configuration (score=\textbf{5.98}), we observe that when retrieving documents using this strategy, the search space for the retriever is limited to a set of context-free chunks. Consequently, the likelihood of retrieving unrelated chunks is slightly higher, as there is no additional contextual information guiding the retriever's decisions. In such cases, the domain-specific cross-encoder reranker, SecureBERT 2.0 (score=\textbf{5.98}) outperforms the LLM reranker (score=\textbf{5.78}). Switching the retrieval method from hybrid to an embeddings-only approach yields similar results.

\subsection{Code Reasoning}

The results of the analysis performed in Section \ref{h1r:fa} are also reflected here. The Contextual Chunking + Hybrid Retrieval + LLM Reranker (score=\textbf{7.10}) and Flat Chunking + Hybrid Retrieval + Cross Encoder Reranker (score=\textbf{7.32}) outperform all other configurations leading to high Code Reasoning Quality for the Falcon H1R model.

\subsection{Remediation Quality}

The trend visualized in Section  \ref{h1r:fa} is clearly visible here as well. The Contextual Chunking + Hybrid Retrieval + LLM Reranker (score=\textbf{6.98}) and Flat Chunking + Hybrid Retrieval + Cross Encoder Reranker (score=\textbf{6.80}) are the top-performing RAG configurations for this criterion.  

\subsection{CWE Remediation Statistics}
To evaluate the quality of the fixes generated by each model using our proposed agentic system, we first select the best configurations for each Falcon model, as presented in Table \ref{tab:rag_comparison_final}. For each selected configuration, we calculate the proportion of CWEs for which a valid fix was generated. This assessment is conducted by examining the logs produced by the Judge Agent. For each test case, the Judge Agent returns an evaluation in accordance with a predefined template

\begin{lstlisting}[language=json]
{
  "structural_integrity": { /* omitted */ },
  "factual_grounding": { /* omitted */ },
  "code_reasoning_quality": { /* omitted */ },
  "remediation_quality": {
    "score": 0,
    "justification": "2-3 sentence explanation.",
    "fix_addresses_root_cause": false,
    "syntax_valid": false
  },
  "overall_score": 0.0
}
\end{lstlisting}

\begin{figure}[h!] 
    \centering
    \includegraphics[width=\linewidth]{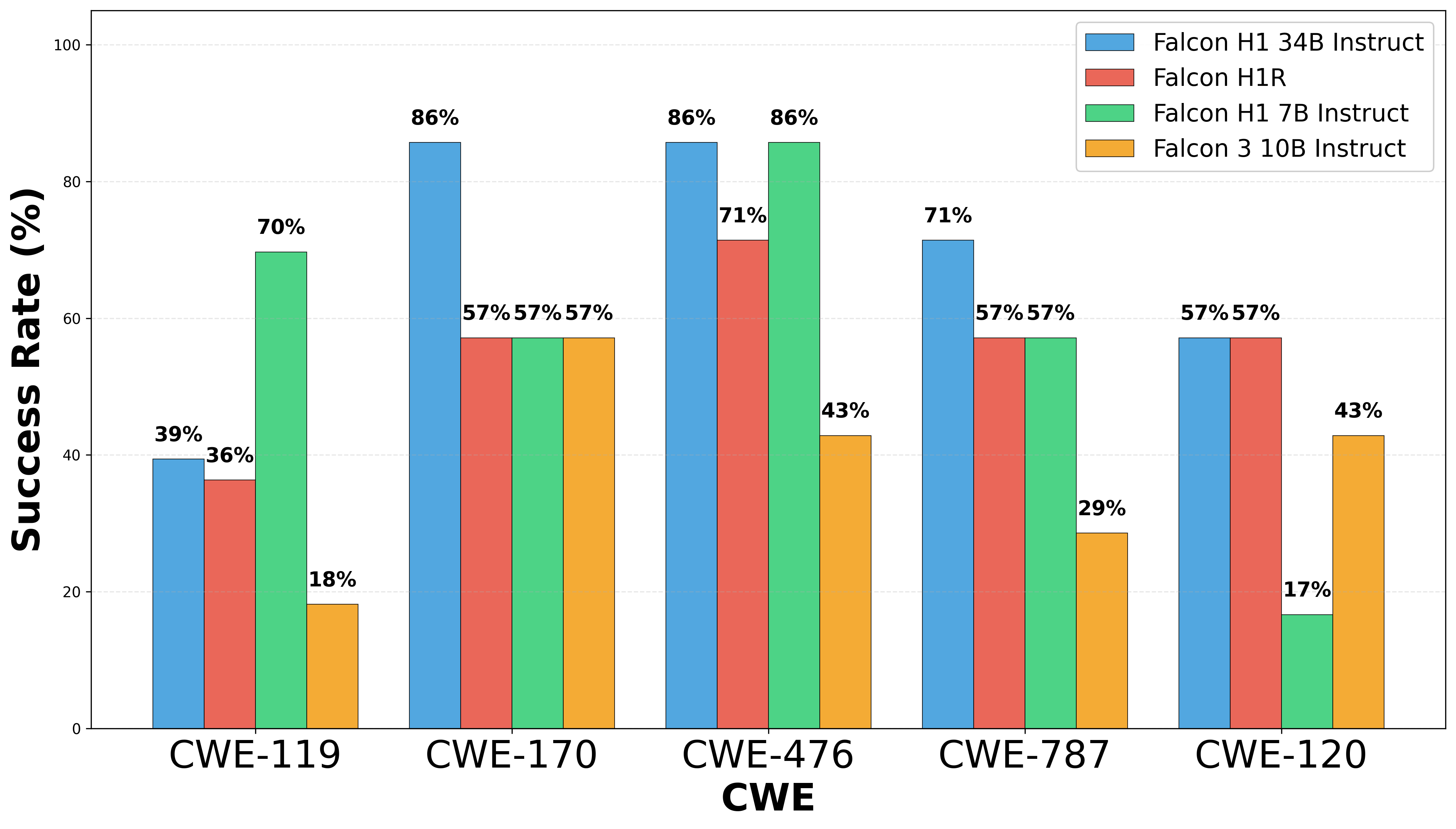} 
    \caption{CWE Remediation Analysis Plot}
    \label{fig:analysis_count}  
\end{figure}

Under the remediation quality criterion, we define two evaluation keys, namely \texttt{fix\_addresses\_root\_cause} and \texttt{syntax\_valid}. These indicate whether the generated fix addresses the root cause specified in the ground-truth annotation and whether the fix is syntactically valid, respectively. A remediation is considered successful only if both conditions are satisfied for both judges. Figure \ref{fig:analysis_count} shows the success ratios for the chosen configs on various CWEs. On average, RAVEN's agentic workflow yields a 54.21\% remediation success rate.

Upon examining the Judge Agent logs for Falcon H1R-7B and Falcon H1-34B-Instruct, we can observe that
\begin{itemize}
    \item Falcon H1R-7B:  This model consistently generates syntactically correct fixes; however, these fixes often prevent program failure by \textbf{\textit{disabling or bypassing the problematic behavior}} rather than directly addressing and correcting the underlying cause.
    
    \item Falcon H1-34B-Instruct: This model reliably generates syntactically valid code, but \textbf{\textit{remediation quality degrades when fixes require understanding intent, control flow, or multi-step reasoning}}.

\end{itemize}

\section{Conclusion}
In this paper, we presented RAVEN, a multi agent framework that combines LLM agents with Retrieval Augmented Generation to produce structured, professional style vulnerability analysis reports from vulnerable source code. By coordinating an Explorer, a retrieval module grounded in curated vulnerability knowledge, an Analyst, and a Reporter, RAVEN supports end to end reasoning from vulnerability identification to impact assessment and remediation guidance. We also introduced a task specific LLM as a Judge to evaluate report quality across structure, alignment with ground truth, code reasoning, and remediation. Experiments on NIST-SARD across diverse CWE categories demonstrate that RAVEN can generate coherent and technically meaningful reports, indicating its promise for scaling vulnerability documentation workflows and assisting human analysts.

\section*{Acknowledgements}

This research was partially funded by Technology Innovation Institute (TII) under the “CASTLE: Cross-Layer Security for Machine Learning Systems IoT” project. Experiments are performed with NYUAD Jubail High Performance Computing.

\bibliographystyle{IEEEtran}
\bibliography{reference}

\end{document}